\documentstyle[12pt]{article}

\begin{document}
\title{Narayana numbers  in  "explicit sufficient  invariants  for an interacting particle system   ( by  Itoh, Mallows, and Shepp)"}
\author{Yoshiaki Itoh \thanks{
itoh@ism.ac.jp}\\
Institute of Statistical Mathematics\\ and Graduate University for Advanced Studies\\ Tachikawa, Tokyo  
} %\\
\maketitle
\begin{abstract}
We consider an interacting particle system on star graphs. As in the case of the Kdv equation, 
we have infinitely many invariants ( here, martingale invariants). 
It  enables
us to obtain the limiting distribution of the Markov chain. 
Each of the martingale invariants is 
 a homogeneous polynomial with coefficients of Narayana numbers.
 The identity for the enumeration of plane unlabeled trees, which gives Narayana numbers,   
becomes the key identity to obtain the probability of death states by a change of variables.
\end{abstract}

\section{Interacting particle system   on graphs }
The KdV equation was shown to have infinitely many invariants by Gardener, 
Green, Kruskal,
Lax, Miura, and Zabusky, see  \cite{ar} p.~543 or \cite{mjd}.
In the interacting particle system on graphs  of Itoh, Mallows and Shepp \cite{ims}, 
the notion of ``invariant'' is replaced by ``martingale''. 
Having a sufficient set of martingale ``invariants''
\cite{ims}  is what enables
us to obtain the limiting distribution of the Markov chain. 

The Narayana number has various combinatorial interpretations,   ballot problem, \cite{bm,  na}, enumeration of plane unlabeled tree \cite{ fs},  Eulerian recurrence \cite{hcd}  etc.  
Here, we point out that each of the martingale invariants \cite{ims} is 
 a homogeneous polynomial with coefficients of Narayana numbers.
We show that a change of variables makes the bivariate generating function for the enumeration of plane unlabeled tree \cite{ fs} the key identity in \cite{ims}.   
 Then we  have  the fundamental  equation in \cite{ims} to obtain the limiting distribution of  
the death states of our previous study \cite{ims}. 

"Suppose initially there are $N_i (0)$ particles at each vertex $i$ of $G$, 
and that the particles interact to form a Markov chain:
at each instant two particles are chosen at random, and if these are at {\em 
adjacent} vertices of $G$, one particle jumps to the other particle's vertex, 
each with probability 1/2.
The process ${\bf N}$ enters a death state after a finite time when all the 
particles are in some {\em independent} subset
of the vertices of $G$, i.e., a set of vertices with no edges between any two 
of them.
The problem is to find the distribution of the death state, $\eta _i= N_i 
( \infty )$, as a function of $N_i (0)$.

We are able to obtain, for some special graphs, the {\em limiting}
distribution of $N_i$ if the total number of particles $N \to \infty$ in such 
a way that the fraction, 
$N_i (0) /N$, at each vertex is held fixed as $N \to \infty$.
In particular we can obtain the limit law for the graph,
$S_2: \cdot$---$\cdot$---$\cdot$, having 3 vertices and 2 edges." (from \cite{ims}).

For the complete graph, the model is that of Moran for the 
Fisher-Wright random sampling effect in population genetics.
In the more general case the model might be applied to study speciation in
biology as well as political positionings \cite{ims,krl,pg,tiya}.

Although the discrete version of the process is thus not amenable to explicit 
calculation, the situation is different for the continuous version obtained by 
letting $N \to \infty$ in such a way for each $i \in G$, that 
$N_i (0) /N $ is held fixed.
The following Ito process is then obtained.

Let $G$ be any undirected simple graph. 
We  define $X_i (t)$, $i \in G$, $t \ge 0$, with $X_i (0)$, as the weak 
solution to the stochastic Ito equation for $t \ge 0$ \cite{ek, ims, ma},
%\beql{eq103}
\begin{eqnarray}
dX_i = \sum_{j \in {\cal N}_i}\sqrt{X_i X_j} d B_{ij}, ~~~ i \in G
\end{eqnarray}
where ${\cal N}_i$ is the set of neighbors of $i$ in $G$, and 
 $B_{ij}$ are
independent
standard Wiener processes for {\em distinct} pairs $\{ i,j \}$ and with the skew-symmetry property,
$$B_{ji} (t) = - B_{ij} (t) , ~~~ t \ge 0 ~.$$

\section{Infinitely many invariants to obtain the probability law for the star graph $S_2$}
Consider the case of the star
graph, $S_2$,
where there are $2$ leaves, each connected to a central root 0. 
\begin{eqnarray}
dX_{0}(t)\nonumber &=&\sqrt {X_{0}(t)X_1 (t)} dB_{01}(t)+\sqrt {X_0(t)X_2 (t)} dB_{02}(t) \\
dX_{1}(t)\nonumber &=& \sqrt {X_{1}(t)X_0 (t)} dB_{10}(t)\\ 
dX_{2}(t)&=& \sqrt {X_{2}(t)X_0 (t)} dB_{20}(t).\label{s2}
\end{eqnarray}
We have the independent subsets of vertices
 $\{1\}$, $\{0\}$, $\{2\}$, $\{1, 2\}$.
 It is clear that for each $i $, $X_i (t)$ is a martingale and 
there exists a first time $\tau \ge 0$, for which
$\{ i: X_i ( \tau ) > 0 \} =J(\tau )$ is an independent
 subset of $S_2$ and 
$P( \tau < \infty) =1$.
The Ito calculus gives 
 one homogeneous polynomial martingale for each $n \ge 2$ for Eq. \ref{s2}, 
 
 The multiplication table for Ito calculus,
McKean (1969) p.~44, 
gives
$$
(dX_i)^2  = X_i \sum_{j \in {\cal N}_i} X_j dt $$
$$(dX_i) (dX_j)  = 
\left\{
\begin{array}{cl}
- X_i X_j dt & j \in {\cal N}_i \\
0 & j \not\in {\cal N}_i ~.
\end{array}
\right.
$$

  If $n\geq 2$ and $Y(t)$ is a homogeneous 
polynomial of degree $n$ in $X_i (t)$, $i \in G$, i.e. a sum of terms of the 
form $\prod\limits_{i \in G} X_i^{a(i)}$, where $a(i)$ are nonnegative integer 
exponents
with $\sum a (i) = n$, then $dY(t) = Q dB + R dt$, where $R$ is again
a homogeneous polynomial of the same degree $n$, 
and $Q$ is the Brownian term.

We have  a set of homogeneous polynomial martingales for the star 
graph, $S_r$, with $r+1$ vertices,
where there are $r$ leaves each connected to a central root, 0 \cite{ims}.
 The simplest case of this is $r=2$, and in this case we can use the
resulting family of martingales to determine the limiting
distribution explicitly \cite{ims}.

Apply the Ito calculus
\begin{eqnarray*}
&& R=\frac{1}{2}\sum_{i=2}^{n-1} \frac{n!}{(n-i)!(i-1)!}\,\,
\frac{(n-2)!}{(n-i-1)! (i-2)!}(-1)^iX_1^{i-1}X_2^{n-i}X_0\\
&+&\frac{1}{2}\sum_{i=1}^{n-2} \frac{n!}{(n-i-1)!i!}\,\,
\frac{(n-2)!}{(n-i-2)! (i-1)!}(-1)^iX_1^{i}X_2^{n-i-1}X_0
\end{eqnarray*}
to have  one
{\bf martingale invariants}  for each $n \ge 2$, 
\begin{eqnarray}
Y_n (t) = \sum_{i=1}^{n-1} {n\choose i}
{n-2 \choose i-1}(-1)^i X_1^i (t) X_2^{n-i} (t).\label{martingale}
\end{eqnarray}
Note $\frac{1}{n}Y_n (t)$ is  a homogeneous polynomial martingale with coefficients of Narayana numbers \cite{ims}.

\section{Generating function of the invariants}
We use $EY_n (\tau) = E_n (0)$
to obtain the laws of $I(\tau)$ and $X( \tau )$ as a function of
$  (X_1(0),X_2(0),X_3(0))$.

The following {\bf key identity }(Itoh, Mallows and Shepp \cite{ims}) is valid for $|v| < 1/4$, $0 \le x \le 1$,
\begin{eqnarray}
\sum_{n \geq 2} \frac{v^n}{n} \sum_{i=1}^{n-1} {n\choose i} {n-2 \choose i-1}
(-1)^i x^i (1-x)^{n-i}
= xv + \frac{1-v}{2}
(1 - \sqrt{ 1 + \frac{4xv}{(1-v)^2}}).\label{mallows}
\end{eqnarray}
Define the generating function for any $u$ the process  with Eq. (\ref{martingale}), 
\begin{eqnarray}
Z_u (t) = \sum_{n \ge 2} \frac{u^n}{n} Y_n (t) ,~~~t \ge 0~.\label{zmartingale}
\end{eqnarray}
$Z_u(t)$ is also a bounded martingale for $|v| < 1/4$. 
Considering $X_1^i ( \tau ) X_2^{n-i} (\tau)=X_1^i ( \tau ) (1-X_1(\tau))^{n-i} (\tau)$, we have from Eq. (\ref{mallows}),
\begin{eqnarray}
 EZ_u ( \tau )= \int_0^1 ( xu + \frac{1-u}{2} ( 1- \sqrt{1+ 
\frac{4xu}{(1-u)^2}} ) ) \mu (dx).\label{tau}
\end{eqnarray}
Put $v=u(X_1(0)+X_2(0))$ and $x=X_1(0)/(X_1(0)+X_2(0))$ to Eq. (\ref{mallows}),
we have from Eq. (\ref{mallows})
\begin{eqnarray}
 Z_u ( 0)= u X_{1}(0) + \frac{1-u(X_1(0)+X_2(0))}{2} ( 1- \sqrt{1+ 
\frac{4uX_1(0)}{(1-u(X_1(0)+X_2(0)))^2}} ).\label{zero}
\end{eqnarray}
For Eq. (\ref{tau}) and Eq. (\ref{zero}), we have the 
{\bf fundamental equation } (Itoh, Mallows, Shepp \cite{ims})
\begin{eqnarray}
EZ_u (\tau) = 
Z_u ( 0) 
\end{eqnarray}\label{fundamental}
for $|v| < 1/4$. 
\vspace{0.2cm}
By using  Eq. (\ref{fundamental}), we obtain 
$\mu (dx) = P(X_1 ( \tau ) \in dx)$ (\cite{ims}), as 
shown in \cite{ims}. 

\section{  Enumeration of plane unlabeled trees for the Narayana numbers}
The identity for the enumeration of plane unlabeled trees $G_{n,i}$   
with $n$ nodes and $i$ leaves of Flajolet and Sedgewick ( page 182 \cite{fs}) 
gives the key identity Eq. (\ref{mallows}) by the change of variables.  
\begin{equation}
G(s,w)=w s+ \sum_{n \ge 2} w^{n} \sum^{n-1}_{i=1} G_{n,i}  s^{i}.\label{series}
\end{equation}
We have 
\begin{equation}
 G(s,w) = w s + w(G(s,w)+ G(s,w)^2 + G(s,w) ^3+\cdots ).\label{tree0}
\end{equation}
Hence we have
 \begin{equation}
 G(s,w) =  w (s + \frac{G(s,w)}{1-G(s,w)}),  \label{tree}
\end{equation}
which is  solved explicitly as
\begin{equation}
 G(s,w) = \frac{1}{2} \left( 1 + (s-1)w - \sqrt{1-2(s+1)w +
(s-1)^{2} w^{2}} \right). \label{quad}
\end{equation}
By using the following Lagrange inversion, from Eq. (\ref{tree}),
we have the Narayana numbers   (\cite{bm,fs,na}) for Eq. (\ref{series}), 
\begin{eqnarray*}
 G_{n,i} &=& [s^{i}] [w^{n}] G(s, w) =  [s^{i}] \frac{1}{n} [y^{n-1}] 
(s + \frac{y}{1-y} )^{n}\\
&=&[s^{i}] \frac{1}{n} [y^{n-1}] 
(s + \frac{y}{1-y} )^{n}\\
&=&  \frac{1}{n} {n \choose i} [y^{n-1}] \frac{y^{n-i}}{(1-y)^{n-i}} 
=\frac{1}{n} {n \choose i} {{n-2} \choose {i-1}} . 
\end{eqnarray*}
where $ [s^{i}][w^{n}] Gsz,w)$ is the coefficient of $s^i w^n$ of $ G(s,w)$ 
and 
 $[y^{n-1}] \frac{y^{n-i}}{(1-y)^{n-i}}$ is the coefficient of 
 $y^{n-1}$ of $ \frac{y^{n-i}}{(1-y)^{n-i}}$. 

%\newpage 
\noindent

{\bf  Lagrange Inversion Theorem} ~~{\it  Let $\phi(u) =
\sum^{\infty}_{j=0} ~\phi_{j} u^{j}$ be a formal power series with
$\phi_{0} \neq 0$, and let $Y(z)$ be the unique formal power series
solution of the equation $Y ~=~ z \phi (Y)$. ~The coefficients of $Y$, 
is  given by}
$$[z^{n}]Y(z)  =\frac{1}{n} [u^{n-1}] (\phi(u))^{n} $$
\vspace{0.2cm}

By using Eq. (\ref{tree}), Lagrange inversion theorem\cite{fs} shows 
the Narayana numbers  $G_{n,i}=\frac{1}{n} {n \choose i} {{n-2} \choose {i-1}}$.  Taking$w=v(1-x)$ and  $s = \frac{-x}{1-x}$ for the above Eq.(\ref{series}) and     Eq. (\ref{quad}),
 we have 
 Eq. (\ref{mallows}), which is the key identity Eq. (\ref{fundamental}) to show 
Eq. (\ref{fundamental}).

{\bf Acknowledgement.} The author greatly thanks Philippe Flajolet for showing
him the Narayana numbers for the enumeration of plane unlabeled trees \cite{fs} in
1999.

\end{document}